\documentclass{article}

\usepackage{arxiv}
\usepackage[square,numbers]{natbib}
\usepackage{graphicx}
\usepackage{amsmath}

\usepackage[utf8]{inputenc} % allow utf-8 input
\usepackage[T1]{fontenc}    % use 8-bit T1 fonts
\usepackage{hyperref}       % hyperlinks
\usepackage{url}            % simple URL typesetting
\usepackage{booktabs}       % professional-quality tables
\usepackage{amsfonts}       % blackboard math symbols
\usepackage{nicefrac}       % compact symbols for 1/2, etc.
\usepackage{microtype}      % microtypography
\usepackage{lipsum}
\usepackage{graphicx}
\graphicspath{ {./images/} }

\title{Exact matrix model for generalized Ising model}

\author{
 Vadym Sakhno
   \And
 Mykola Sakhno
}

\begin{document}
\maketitle
\begin{abstract}
The Ising model was generalized to a system of cells interacting exclusively by presence of shared spins. Within the cells there are interactions of any complexity, the simplest intracell interactions come down to the Ising model. The system may be not only one-dimensional but also two-dimensional, three-dimensional, etc.

The purpose of the paper is to develop an approach to constructing the exact matrix model for any considered system in the simplest way. Without this, it is almost impossible to analyze complex systems. After trying a lot of ways, the approach has been simplified to suit first year students.

Using the approach, the exact matrix model for a two-dimensional generalized Ising model was constructed.

The following approaches have already been developed to obtaining and analyzing partition function and various thermodynamic functions. They will be considered in following papers.
\end{abstract}

\section{Introduction} \label{sec:Introduction}

\subsection{Microreview} \label{subsec:Microreview}
The Ising model arouses great interest in various fields. And even in comparatively narrow field of the Ising model exact solutions, there are thousands of publications. The authors have tried to develop their own approaches different from the published ones in order to get results worthy of Your attention. Therefore, the authors limited themselves to a microreview of only those works that are most directly related to their approaches.
As foundation base of statistical mechanics, the authors use K. Huang \cite{huang1987statistical}, of Ising model exact solutions – L. Onsager \cite{onsager1944crystal}, as fundamental review of exactly solved models – R.J. Baxter \cite{baxter2016exactly}. Among the numerous publications developing exact solutions, the closest seems to be P. Khrapov \cite{khrapov2020disorder} and Yu.D. Panov \cite{panov2020local}. And research directions were outlined by B.M. McCoy and J.-M. Maillard \cite{mccoy2012importance}. 

\subsection{Systems and tasks under consideration} \label{subsec:SystemsAndTasksUnderConsideration}
This paper considers systems depending on variables with a finite range of values (hereinafter, for brevity, each such variable is called spin, and the amount of its values - magnitude). These systems may have arbitrary: physical nature, the dimensions of space-time, geometry, other properties; but must have three given properties:

\begin{enumerate}
  \item System energy E can be represented as the sum of energies of its parts called cells. 
  \item Each cell energy is the given function only of its own spins as well as internal and external parameters. Interaction between cells is performed exclusively through existence of shared spins. There are no other interactions between cells. The structure and parameters of different cells may vary.
  \item There is such cell numbering that each cell has at least one spin belonging exclusively to cells with a lesser number. 
\end{enumerate}

For these systems, the following tasks are considered:

\begin{enumerate}
    \item Construct the exact matrix model.
    \item Using the exact matrix model, find the partition function exactly and analyse it
    \begin{equation} \label{eq:1}
        Z=\sum exp(-\frac{E}{k_B*T}),
    \end{equation}
    where summing is performed over all possible values of all spins.
    \item Using the partitioning function (\ref{eq:1}), find the thermodynamic functions exactly and analyze them: Helmholtz free energy A, magnetization M, magnetic susceptibility $\chi$, etc.
    \begin{equation}
        A=-k_B*T*ln(Z), \ M=-\frac{\partial}{\partial H}(\frac{A}{k_B * T}), \ \chi=\frac{\partial M}{\partial H}.
    \end{equation}
\end{enumerate}

This paper considers only the first task, the rest ones will be discussed in following publications.

\subsection{Notes} \label{subsec:Notes}

\begin{enumerate}
    \item For d-magnitude spins, we assume that it has d values from \( (-\frac{d-1}{2}) \) to \( (+\frac{d-1}{2})\), with step 1. For example, 2-magnitude spin has two values \( \{-\frac{1}{2},\frac{1}{2}\} \), 3-magnitude spin has 3 values \( \{-1,\ 0,\ 1\} \), etc.
    \item The system may be split into cells by various ways. For example, some cells may be merged, and some other sell may be split. A shared part of two cells may be included in the first cell or the second cell or split in some way between these two cells. Such transformations do not influence the final result; however, they may simplify calculations.
    \item The Ising model is obtained using the simplest energy function for cells. Thus, the considered systems are generalizations of the Ising model. For example, 4-spin cell considers not only interactions of its 4 spins with an external field and spin-spin interactions between nearest spins but also other spin-spin interactions as well as 3-spin and 4-spin interactions.
    \item Typical cell numbering is suitable for typical systems. 
\end{enumerate}

\subsection{Paper overview} \label{subsec:PaperOverview}

Simple approach to construction of the exact matrix model for any considered system is developed in section \ref{sec:ConstructionOfExactMatrix}.

A simplified general example of this construction is shown in section \ref{sec:GeneralSimplifiedExample}.

A demonstration section was written initially. The exact matrix models were constructed for various one-dimensional systems there. But the authors got a comment on lack of novelty. Thus, the section was deleted.

The exact matrix model for a two-dimensional system is constructed in section \ref{sec:TwoDimensionalSystem}.

\section{A simple approach to construction of the exact matrix model for any considered system} \label{sec:ConstructionOfExactMatrix}

The proposed approach consists of three simple steps: the first two steps are construction of two intermediate models and then a sought-for matrix model.

\subsection{Construction of intermediate function model}

Let the system consists of \( N+2 \) cells; each cell is assigned a unique number \( n\in[0, \ N+1] \). The cell numbered 0 is called the start cell, the cell numbered \( N+1 \) is called the finish cell, the remaining cells numbered \( n\in[1,\ N] \) are called internal cells.

A spin belonging to only one cell is called the local spin. The set of all local spins of the cell n is denoted by \( Y_n \). Any local spin \( y_{n\nu} \) of this set is numbered with a unique compound number consisted of two sub-numbers: n – cell number,  $\nu$ – serial number inside the cell. The set of all local spins of the system is denoted as \( Y=Y_0 \cup Y_1 \cup Y_2 \cup \ldots \cup Y_N \cup Y_{N+1} \). Some local spins may have infinite range of values and even be continuous.

A spin belonging to two or more cells is called the shared spin. Any shared spin \( x_{n\nu} \) is numbered with a unique compound number consisted of two sub-numbers: n – the maximum number of the cell containing the spin, $\nu$ – serial number in shared spins having the first sub-number n. The set of all shared spins belonging to cell (not only having first sub-number n) is denoted by \( X_n \). The set of all shared spins of the system is denoted as \( X=X_0 \cup X_1 \cup X_2 \cup \ldots \cup X_N \cup X_{N+1} \).

The energy of cell n is denoted as \( E_n(X_n,\ Y_n) \).

Let us sum up in (\ref{eq:1}) over all local spins. Summation over continuous spin becomes integration. Of course, cells without local spins do not need to be summed over them. Thus, for each cell \( n \in [0, \ N+1 ] \) we introduce cell function \( Z_n(X_n) \)

\begin{equation} \label{eq:3}
        Z_n(X_n)=\sum_{Y_n} exp \left( -\frac{E_n(X_n,\ Y_n)}{k_B T} \right)>0.
\end{equation}

Property 3 in subsection \ref{subsec:SystemsAndTasksUnderConsideration} implies that for each cell numbered \( n \in [1, \ N+1] \) there is a non-empty subset \( \Upsilon_n \subset X_n \) of shared spins with the first sub-number n. Then summation over shared spins in (\ref{eq:1}) may be split into (N+1) partial summations 

\begin{equation} \label{eq:4}
    Z=\sum_{\Upsilon_{N+1}}{Z_{N+1}\left(X_{N+1}\right)}\ast\sum_{\Upsilon_N}{Z_N\left(X_N\right)\ast}\ldots\ast\sum_{\Upsilon_n}{Z_n\left(X_n\right)}\ast\ldots\ast\sum_{\Upsilon_2}{Z_2\left(X_2\right)}\ast\sum_{\Upsilon_1}{Z_1\left(X_1\right)\ast Z_0\left(X_0\right)},
\end{equation}

where each partial summation \( n \in [1, \ N+1] \) is performed over values of shared spins of set \( \Upsilon_n \).

The equation (\ref{eq:4}) uses the cell functions, so we call (\ref{eq:4}) the function model.

\subsection{Construction of intermediate frame model}

The function model (\ref{eq:4}) uses each cell function only in the values of its shared spins. Thus, each cell function generates a set of its values. Let us call the set a cell frame and enumerate its values as follows. 

For each \( d_{n\nu} \) - magnitude shared spin \( x_{n\nu} \in [-\frac{d_{n\nu}-1}{2}, \ +\frac{d_{n\nu}-1}{2}] \) let us introduce a corresponding spin-number \( i_{n\nu} \) like this

\begin{equation} \label{eq:5}
    i_{n\nu}=x_{n\nu}+\frac{d_{n\nu}-1}{2}.
\end{equation}

The spin-number \( i_{n\nu} \in [0, \ d_{n\nu}-1] \) is used for numbering values in any cell frame containing the spin. 

The inverse equation for (\ref{eq:5}) is

\begin{equation} \label{eq:6}
    x_{n\nu}=i_{n\nu}-\frac{d_{n\nu}-1}{2}.
\end{equation}

This results in a simple algorithm of constructing the frame for each cell function \( Z_n \left( X_n \right) \). In \( Z_n \left( X_n \right) \) each spin of \( X_n \) is substituted with its spin-number using (\ref{eq:6}). Let us denote the set of these spin-numbers as \( I_n \) and use it for numbering in the cell frame. Then the cell frame \( Z_{nI_n} \) for the cell function \( Z_n\left(X_n\right) \) is

\begin{equation} \label{eq:7}
    Z_{nI_n}=Z_n \left( X_n \left( I_n \right) \right) > 0.
\end{equation}

Example. Let a cell function \( Z_n \) have three 2-magnitude spins \(Z_n\left(x_{n+1},x_n,x_{n-1}\right)\). The cell frame \( Z_{ni_{n+1}i_ni_{n-1}}=Z_n\left(i_{n+1}-\frac{1}{2}, \ i_n-\frac{1}{2}, \ i_{n-1}-\frac{1}{2}\right) \) consists of eight elements, for which spin-numbers \( i_{n+1},i_n,i_{n-1} \) have two values \( \left\{0,\ 1\right\} \) each. Namely: \( Z_{n000}=Z_n\left(-\frac{1}{2},-\frac{1}{2},-\frac{1}{2}\right) \), \( Z_{n001}=Z_n\left(-\frac{1}{2},-\frac{1}{2},\frac{1}{2}\right) \), \( Z_{n010}=Z_n\left(-\frac{1}{2},\frac{1}{2},-\frac{1}{2}\right) \), \( Z_{n011}=Z_n\left(-\frac{1}{2},\frac{1}{2},\frac{1}{2}\right) \), \( Z_{n100}=Z_n\left(\frac{1}{2},-\frac{1}{2},-\frac{1}{2}\right) \), \( Z_{n101}=Z_n\left(\frac{1}{2},-\frac{1}{2},\frac{1}{2}\right) \), \( Z_{n110}=Z_n\left(\frac{1}{2},\frac{1}{2},-\frac{1}{2}\right) \), \( Z_{n111}=Z_n\left(\frac{1}{2},\frac{1}{2},\frac{1}{2}\right) \). Thus, set of spin-numbers  \( i_{n+1},i_n,i_{n-1} \) may be considered as a binary number having values from 0 to 7.

In the function model (\ref{eq:4}), substituting each spin with its spin-number, and each cell function with its frame constructs the frame model

\begin{equation} \label{eq:8}
    Z=\sum_{I(\Upsilon_{N+1})}Z_{(N+1)I_{N+1}}\ast\sum_{I\left(\Upsilon_N\right)}Z_{NI_N}\ast\ldots\ast\sum_{I(\Upsilon_n)}Z_{nI_n}\ast\ldots\ast\sum_{I(\Upsilon_2)}Z_{2I_2}\ast\sum_{I(\Upsilon_1)}{Z_{1I_1} \ast Z_{0I_0}},
\end{equation}

where \( I(\Upsilon_n) \) with \( n \in \left[1, \ N+1\right] \) – spin-numbers set for spins of set \( \Upsilon_n \).

The constructing of the frame model may be substantiated more formally. In the function model (\ref{eq:4}), let us interpolate each cell function in the Lagrange form. Then the coefficients of Lagrange basis polynomials are the frame values. Taking into account the orthonormality of the Lagrange basis polynomials, we obtain the frame model (\ref{eq:8}).

\subsection{Construction of the sought matrix model}

When in the frame model (\ref{eq:8}) each cell frame is rewritten in the matrix form, then the summations turn into matrix multiplications. To do this, in the frame model (\ref{eq:8}) perform the following steps:
\begin{enumerate}
    \setcounter{enumi}{-1}
    \item The frame of the start cell \( Z_{0I_0} \) is rewritten as a column vector \( \overleftarrow{Z_0} \).
    \item Denote the result of the first summation as a column vector \( \overleftarrow{A_1} \). Denote the frame of the first cell \( Z_{1I_1} \) as a matrix \( \overleftrightarrow{Z_1} \) so that \( \overleftarrow{A_1}=\overleftrightarrow{Z_1}*\overleftarrow{Z_0} \).
    \item Denote the result of the second summation as a column vector \( \overleftarrow{A_2} \). Denote the frame of the second cell \( Z_{2I_2} \) as a matrix \( \overleftrightarrow{Z_2} \) so that \( \overleftarrow{A_2}=\overleftrightarrow{Z_2}*\overleftarrow{A_1} \).
    \item[...]
    \item[N.] Denote the result of N-th summation as a column vector \( \overleftarrow{A_N} \). Denote the frame of the N-th cell \( Z_{NI_N} \) as a matrix \( \overleftrightarrow{Z_N} \) so that \( \overleftarrow{A_N}=\overleftrightarrow{Z_N} * \overleftarrow{A_{N-1}} \).
    \item[N+1.] The frame of the finish cell \( Z_{(N+1)I_{N+1}} \) is rewritten as a row vector \( \overrightarrow{Z_{N+1}} \). Then the resulting partitioning function is \( Z=\overrightarrow{Z_{N+1}} * \overleftarrow{A_N} \).
\end{enumerate}

Combining all these steps, the sought-for matrix model is obtained
\begin{equation} \label{eq:9}
    Z=\overrightarrow{Z_{N+1}} * \overleftrightarrow{Z_N} * \ldots *\overleftrightarrow{Z_n} * \ldots * \overleftrightarrow{Z_2} * \overleftrightarrow{Z_1}* \overleftarrow{Z_0}=\overrightarrow{Z_{N+1}}* \left( \prod_{n=N}^1 \overleftrightarrow{Z_n} \right) * \overleftarrow{Z_0}.
\end{equation}

\section{A general simplified example of obtaining the exact matrix model, and its analysis} \label{sec:GeneralSimplifiedExample}

In the function model (\ref{eq:4}) let us consider only the first summation from the start cell. The rest of the summations are considered similarly. 
Let cells 0 and 1 each have three 2-magnitude spins, and the first summation is

\begin{equation} \label{eq:10}
    A_1\left(x_{20},x_{30},x_{40}\right) = \sum_{x_{10} \in \{ -\frac{1}{2},\frac{1}{2} \} } {Z_1 \left(x_{10},x_{20},x_{40}\right)\ast Z_0\left(x_{10},x_{20},x_{30}\right)}.
\end{equation}

\subsection{Spin types with respect to summation}

For spins of start cell function \( Z_0 \) let us introduce the following types: 
\begin{enumerate}
    \item Spins belonging to the start and first cells exclusively and not belonging to any other cell. The summation “annihilates” them, so they are called “annihilated” spins. Property 3 in subsection \ref{subsec:SystemsAndTasksUnderConsideration} requires “annihilated” spins to be present. There is one “annihilated” spin \(x_{10} \) in (\ref{eq:10}).
    \item Spins belonging to the start, first, and at least one more cell. The summation “processes” them and passes them further, so they are called “processed” spins. “Processed” spins may be missing. There is one “processed” spin \( x_{20} \) in (\ref{eq:10}).
    \item Spins not belonging to the first, but belonging to the start and at least one more cell. The summation just “passes” them further, so they are called “passed” spins. “Passed” spins may be missing. There is one “passed” spin \( x_{30} \) in (\ref{eq:10}).
\end{enumerate}

The first cell has “destroyed” and “processed” spins shared with the start cell. In addition, the first cell may have spins of one more type: 

\begin{enumerate}
    \setcounter{enumi}{3}
    \item Spins not belonging to the start, but belonging to the first and at least one more cell. The summation “creates” them, they are called “created” spins. There is one “created” spin \( x_{40} \) in (\ref{eq:10}).
\end{enumerate}

Thus, example (\ref{eq:10}) is general because it contains spins of all types. Simultaneously, this example is simplified because it contains only one spin of each type, and the spins are 2-magnitude.

\subsection{Construction of the sought matrix model} \label{subsec:ExampleMatrixModel}

First, from the function model (\ref{eq:10}) let us obtain the frame model in the form (\ref{eq:8}). In the function model, we need to substitute each spin with its spin-number, and each function with its frame. Frames of given cell functions \( Z_0\left(x_{10},x_{20},x_{30}\right) \) and \( Z_1\left(x_{10},x_{20},x_{40}\right) \) are calculated using (\ref{eq:7}). Moreover, this calculation is shown in detail in the example after (\ref{eq:7}). The frame of the sought-for function \( A_1\left(x_{20},x_{30},x_{40}\right) \) is just written. Then the frame model of (\ref{eq:10}) is

\begin{equation} \label{eq:11}
    A_{1i_{20}i_{30}i_{40}}=\sum_{ i_{10} \in \left\{ 0,\ 1 \right\} } {Z_{1i_{10}i_{20}i_{40}} \ast Z_{0i_{10}i_{20}i_{30}}}.
\end{equation}

Let’s write (\ref{eq:11}) line by line. In the zero line: \( i_{20}=0,i_{30}=0,i_{40}=0 \); in the first line: \(i_{20}=0,i_{30}=0,i_{40}=1 \); … ; in the seventh line: \(i_{20}=1,i_{30}=1,i_{40}=1 \)

\begin{equation} \label{eq:12}
    \begin{aligned} 
        A_{1000}=Z_{1000} \ast Z_{0000}+Z_{1100} \ast Z_{0100}, \\
        A_{1001}=Z_{1001} \ast Z_{0000}+Z_{1101} \ast Z_{0100}, \\
        A_{1010}=Z_{1000} \ast Z_{0001}+Z_{1100} \ast Z_{0101}, \\
        A_{1011}=Z_{1001} \ast Z_{0001}+Z_{1101} \ast Z_{0101}, \\
        A_{1100}=Z_{1010} \ast Z_{0010}+Z_{1110} \ast Z_{0110}, \\
        A_{1101}=Z_{1011} \ast Z_{0010}+Z_{1111} \ast Z_{0110}, \\
        A_{1110}=Z_{1010} \ast Z_{0011}+Z_{1110} \ast Z_{0111}, \\
        A_{1111}=Z_{1011} \ast Z_{0011}+Z_{1111} \ast Z_{0111}.
    \end{aligned}
\end{equation}

Let us introduce column vectors \( A_1 \) and  \( Z_0 \) and write (\ref{eq:12}) in matrix form

\begin{equation} \label{eq:13}
    \begin{pmatrix}
        A_{1000} \\
        A_{1001} \\
        A_{1010} \\
        A_{1011} \\
        A_{1100} \\
        A_{1101} \\
        A_{1110} \\
        A_{1111}
    \end{pmatrix}
    =
    \begin{pmatrix}
        Z_{1000} & 0 & 0 & 0 & Z_{1100} & 0 & 0 & 0 \\
        Z_{1001} & 0 & 0 & 0 & Z_{1101} & 0 & 0 & 0 \\
        0 & Z_{1000} & 0 & 0 & 0 & Z_{1100} & 0 & 0 \\
        0 & Z_{1001} & 0 & 0 & 0 & Z_{1101} & 0 & 0 \\
        0 & 0 & Z_{1010} & 0 & 0 & 0 & Z_{1110} & 0 \\
        0 & 0 & Z_{1011} & 0 & 0 & 0 & Z_{1111} & 0 \\
        0 & 0 & 0 & Z_{1010} & 0 & 0 & 0 & Z_{1110} \\
        0 & 0 & 0 & Z_{1011} & 0 & 0 & 0 & Z_{1111}
    \end{pmatrix}
    *
    \begin{pmatrix}
        Z_{0000} \\
        Z_{0001} \\
        Z_{0010} \\
        Z_{0011} \\
        Z_{0100} \\
        Z_{0101} \\
        Z_{0110} \\
        Z_{0111}
    \end{pmatrix}
    ,
\end{equation}

or compactly
\begin{equation} \label{eq:14}
    \overleftarrow{A_1} = \overleftrightarrow{Z_1} * \overleftarrow{Z_0}.
\end{equation}

Hence the conclusions on the structure of the matrix \(\overleftrightarrow{Z_1}\) are
\begin{enumerate}
    \item The matrix \( \overleftrightarrow{Z_1} \) is non-negative.
    \item The amount of columns in \( \overleftrightarrow{Z_1} \) is equal to the product of the magnitudes of “annihilated”, “processed” and “passed” spins, that is, all spins of the start cell function (in this example it is 8).
    \item The amount of non-zero columns in \( \overleftrightarrow{Z_1} \) is equal to the product of the magnitudes of “annihilated” spins (in this example it is 2).
    \item The amount of rows in \( \overleftrightarrow{Z_1} \) is equal to the product of the magnitudes of “created”, “processed” and “passed” spins (in this example it is 8).
    \item The amount of non-zero rows in \( \overleftrightarrow{Z_1} \) is equal to the product of the magnitudes of “created” spins (in this example it is 2).
\end{enumerate}

\subsection{The simplest cyclic shift matrices} \label{subsec:SimplestCyclicShiftMatrices}

In (\ref{eq:11}) the start cell frame \( Z_{0i_{10}i_{20}i_{30}} \) contains spin-numbers \( i_{20}i_{30} \). The first sum frame \( A_{1i_{20}i_{30}i_{40}} \) contains the same spin-numbers shifted to the left. Let us introduce cyclic left shift operator \( P_l \), which for any vector \( Z_{i_1i_2i_3} \) performs the transformation

\begin{equation*}
    Z_{i_2i_3i_1}=P_l \ast Z_{i_1i_2i_3}.
\end{equation*}

Let us move on to the matrix form

\begin{equation*}
    \begin{pmatrix}
        Z_{000} \\
        Z_{010} \\
        Z_{100} \\
        Z_{110} \\
        Z_{001} \\
        Z_{011} \\
        Z_{101} \\
        Z_{111}
    \end{pmatrix}
    =
    \begin{pmatrix}
        1 & 0 & 0 & 0 & 0 & 0 & 0 & 0 \\
        0 & 0 & 1 & 0 & 0 & 0 & 0 & 0 \\
        0 & 0 & 0 & 0 & 1 & 0 & 0 & 0 \\
        0 & 0 & 0 & 0 & 0 & 0 & 1 & 0 \\
        0 & 1 & 0 & 0 & 0 & 0 & 0 & 0 \\
        0 & 0 & 0 & 1 & 0 & 0 & 0 & 0 \\
        0 & 0 & 0 & 0 & 0 & 1 & 0 & 0 \\
        0 & 0 & 0 & 0 & 0 & 0 & 0 & 1
    \end{pmatrix}
    *
    \begin{pmatrix}
        Z_{000} \\
        Z_{001} \\
        Z_{010} \\
        Z_{011} \\
        Z_{100} \\
        Z_{101} \\
        Z_{110} \\
        Z_{111}
    \end{pmatrix}
    .
\end{equation*}

Hence the left cyclic shift matrix \( \overleftrightarrow{P_l} \) and its inverse as right cyclic shift matrix \( \overleftrightarrow{P_r}=\overleftrightarrow{P_l}^{-1} \)

\begin{equation} \label{eq:15}
    \overleftrightarrow{P_l}
    =
    \begin{pmatrix}
        1 & 0 & 0 & 0 & 0 & 0 & 0 & 0 \\
        0 & 0 & 1 & 0 & 0 & 0 & 0 & 0 \\
        0 & 0 & 0 & 0 & 1 & 0 & 0 & 0 \\
        0 & 0 & 0 & 0 & 0 & 0 & 1 & 0 \\
        0 & 1 & 0 & 0 & 0 & 0 & 0 & 0 \\
        0 & 0 & 0 & 1 & 0 & 0 & 0 & 0 \\
        0 & 0 & 0 & 0 & 0 & 1 & 0 & 0 \\
        0 & 0 & 0 & 0 & 0 & 0 & 0 & 1
    \end{pmatrix}
    ,\ 
    \overleftrightarrow{P_r}
    =
    \begin{pmatrix}
        1 & 0 & 0 & 0 & 0 & 0 & 0 & 0 \\
        0 & 0 & 0 & 0 & 1 & 0 & 0 & 0 \\
        0 & 1 & 0 & 0 & 0 & 0 & 0 & 0 \\
        0 & 0 & 0 & 0 & 0 & 1 & 0 & 0 \\
        0 & 0 & 1 & 0 & 0 & 0 & 0 & 0 \\
        0 & 0 & 0 & 0 & 0 & 0 & 1 & 0 \\
        0 & 0 & 0 & 1 & 0 & 0 & 0 & 0 \\
        0 & 0 & 0 & 0 & 0 & 0 & 0 & 1
    \end{pmatrix}
    .
\end{equation}

Then matrix \( \overleftrightarrow{Z_1} \) from (\ref{eq:13}) may be written in a form more convenient for analysis

\begin{equation*}
    \overleftrightarrow{Z_l} 
    = 
    \overleftrightarrow{Z_l} * \overleftrightarrow{P_l} * \overleftrightarrow{P_r}
    =
    \begin{pmatrix}
        Z_{1000} & Z_{1100} & 0 & 0 & 0 & 0 & 0 & 0 \\
        Z_{1001} & Z_{1101} & 0 & 0 & 0 & 0 & 0 & 0 \\
        0 & 0 & Z_{1000} & Z_{1100} & 0 & 0 & 0 & 0 \\
        0 & 0 & Z_{1001} & Z_{1101} & 0 & 0 & 0 & 0 \\
        0 & 0 & 0 & 0 & Z_{1010} & Z_{1110} & 0 & 0 \\
        0 & 0 & 0 & 0 & Z_{1011} & Z_{1111} & 0 & 0 \\
        0 & 0 & 0 & 0 & 0 & 0 & Z_{1010} & Z_{1110} \\
        0 & 0 & 0 & 0 & 0 & 0 & Z_{1011} & Z_{1111}
    \end{pmatrix}
    *
    \overleftrightarrow{P_r}
    .
\end{equation*}

\section{A two-dimensional system} \label{sec:TwoDimensionalSystem}

In this paper, we restrict ourselves to considering only one of the two-dimensional systems, a particular case of which is easily diagonalized. The diagonalization will be considered in a following paper.

\subsection{Description of the 2D system under consideration}
\begin{figure} % picture
    \centering
    \includegraphics[width=\textwidth]{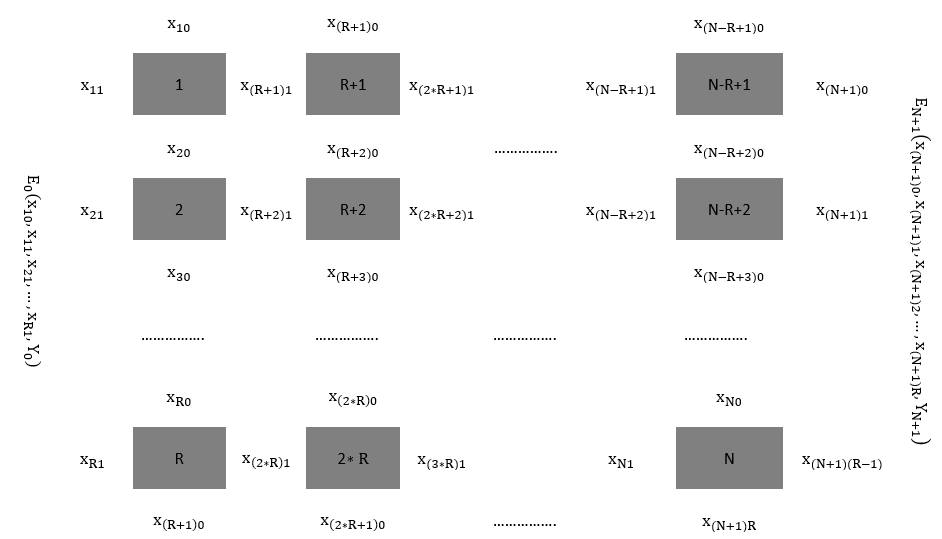}
    \caption{The 2D system under consideration.}
    \label{fig:fig1}
\end{figure}

Let us consider the 2D model shown in Figure \ref{fig:fig1}.

The system consists of \( (N+2) \) cells: $N$ internal cells numbered from $1$ to $N$, start cell numbered $0$, and finish cell numbered \( (N+1) \). 

$N$ internal cells form $R$ rows. The lowest spin of a column continues into the highest spin of the next column (see \( x_{\left(R+1\right)0} \)). Thus, cells are placed along a helix. The last column may be uncompleted. 

Each internal cell $n$ has set $X_n$ of four shared 2-magnitude spins. The numbers of the upper $x_{n0}$ and the left $x_{n1}$ spins have their first sub-number equal to the cell number $n$. Then the lower spin is numbered $x_{(n+1)0}$ as the upper one for cell $(n + 1)$. And the right spin is numbered  $x_{(n+R)2}$ as the left one for cell $(n + R)$. Moreover, each internal cell $n$ may contain an arbitrary set of local spins $Y_n$. The internal cell energy is a given function $E_n\left(X_n,\ Y_n\right)$. Substituting it into (\ref{eq:3}), we get the internal cell function $Z_n\left(x_{n0},\ x_{n1},x_{(n+1)0},x_{(n+R)1}\right)$.

The start and finish cells each contain $(R + 1)$ shared spins and may contain arbitrary sets of local spins. Their energies are given functions $E_0\left(x_{10},\ x_{11},\ x_{21},\ldots,x_{R1},\ Y_0\right)$ and $E_{N+1}\left(x_{(N+1)0},x_{(N+1)1},x_{(N+1)2},\ldots,x_{(N+1)R},\ Y_{N+1}\right)$. Substituting them into (\ref{eq:3}), we obtain the functions of the start and the finish cells $Z_0\left(x_{10},\ x_{11},\ x_{21},\ldots,x_{R1}\right)$ and $Z_{N+1}\left(x_{(N+1)0},x_{(N+1)1},x_{(N+1)2},\ldots,x_{(N+1)R}\right)$.

\subsection{Construction of the matrix model}

% TODO: check section ref
Construction of the matrix model is similar to the general simplified example (section \ref{sec:GeneralSimplifiedExample}).

In the function model (\ref{eq:4}) let us consider only the first summation from the start cell. The rest of the summations are considered similarly.
Let the result function of the first summation be denoted by $A_1$. Then the first summation is

\begin{equation} \label{eq:16}
 A_1
 =
 \sum_{\Upsilon_1}{Z_1\left(X_1\right) \ast Z_0\left(X_0\right)}
 =
    \sum_{\Upsilon_1}{Z_1\left(x_{10},x_{11},\ x_{20},x_{(1+R)1}\right)
    \ast
    Z_0\left(x_{10},\ x_{11},\ x_{21},\ldots,x_{R1}\right)}.
\end{equation}

Set $X_0$ of spins of the start cell function $Z_0$ has spins of two types (see subsection 3.1): subset  of two “annihilated” spins  $\Upsilon_1=\left\{x_{10},\ x_{11}\right\}$ and subset of $(R-1)$ "passed" spins $\left\{x_{21},\ldots,x_{R1}\right\}$. Set $X_1$ of spins of the first cell function $Z_1$ also has spins of two types: the same subset $\Upsilon_1$ of two “annihilated” spins and subset of two “created” spins $\left\{x_{20},x_{(1+R)1}\right\}$. Then the set of spins of the first result function $A_1$ consisting of the "passed" and “created” spins is $\left\{x_{20},\ x_{21}\ ,\ldots,x_{R1},x_{\left(1+R\right)1}\right\}$.

Based on the conclusions from subsection \ref{subsec:ExampleMatrixModel}, we can conclude the sought matrix \(\overleftrightarrow{Z_1}\) structure:
\begin{enumerate}
    \item There are two 2-magnitude “annihilated” spins, so the amount of non-zero columns is $2^2=4$.
    \item There are $(R+1)$ of 2-magnitude “annihilated” and “passed” spins, so the amount of columns is $2^{R+1}$.
    \item There are two 2-magnitude “created” spins, so the amount of non-zero rows is $2^2=4$.
    \item There are $(R+1)$ of 2-magnitude “created” and “passed” spins, so the amount of rows is $2^{R+1}$.
\end{enumerate}

In (\ref{eq:16}), we need to substitute each spin with its spin-number and each function with its frame. Frames of given cell functions $Z_0\left(x_{10},\ x_{11},\ x_{21},\ldots,x_{R1}\right)$ and $Z_1\left(x_{10},x_{11}, x_{20},x_{(1+R)1}\right)$ are calculated using (\ref{eq:7}). The frame of the first result function $A_1\left(x_{20},\ x_{21}\ ,\ldots,x_{R1},x_{\left(1+R\right)1}\right)$ is just written.

Let us describe in detail the frame of the first cell function $Z_1\left(x_{10},x_{11}, x_{20},x_{(1+R)1}\right)$. For each spin \( x_{n\nu} \in [-\frac{1}{2}, \ +\frac{1}{2}] \) we introduce a corresponding spin-number \( i_{\nu} \in [0, 1] \) according to (\ref{eq:6}). Then, according to (\ref{eq:7}), the sought-for first cell frame with 16 values is

\begin{equation} \label{eq:17}
    Z_{1i_{10}i_{11}i_{20}i_{\left(R+1\right)1}}=Z_1\left(i_{10}-\frac{1}{2}, \ i_{11}-\frac{1}{2}, \ i_{20}-\frac{1}{2}, i_{\left(R+1\right)1}-\frac{1}{2}\right).
\end{equation}

After the described substitution of functions with frames and spins with spin-numbers in the function model (\ref{eq:16}), we obtain the frame model

\begin{equation} \label{eq:18}
    A_{1i_{20}i_{21}\ldots i_{R1}i_{\left(1+R\right)1}}
    =
    \sum_{i_{10} \in \left\{0,\ 1\right\}\ \ \ i_{11}\in\left\{0,\ 1\right\}}
        {
            Z_{1i_{10}i_{11}i_{20}i_{(1+R)1}}
            *
            Z_{0i_{10}i_{11}i_{21}\ldots i_{R1}}
        }.
\end{equation}

Let us write (\ref{eq:18}) line by line
\begin{equation} \label{eq:19}
    \begin{aligned} 
        A_{100 \ldots 00} = Z_{10000} \ast Z_{0000 \ldots 0} + Z_{10100} \ast Z_{0010 \ldots 0} + Z_{11000} \ast Z_{0100 \ldots 0}+Z_{11100} \ast Z_{0110 \ldots 0}, \\
        A_{100 \ldots 01} = Z_{10001} \ast Z_{0000 \ldots 0} + Z_{10101} \ast Z_{0010 \ldots 0} + Z_{11001} \ast Z_{0100 \ldots 0} + Z_{11101} \ast Z_{0110 \ldots 0}, \\
        A_{100 \ldots 10} = Z_{10000} \ast Z_{0000 \ldots 1} + Z_{10100} \ast Z_{0010 \ldots 1} + Z_{11000} \ast Z_{0100 \ldots 1} + Z_{11100} \ast Z_{0110 \ldots 1}, \\
        A_{100 \ldots 11} = Z_{10001} \ast Z_{0000 \ldots 1} + Z_{10101} \ast Z_{0010 \ldots 1} + Z_{11001} \ast Z_{0100 \ldots 1} + Z_{11101} \ast Z_{0110 \ldots 1}, \\
        \dots \\
        A_{101 \ldots 00} = Z_{10000} \ast Z_{0001 \ldots 0} + Z_{10100} \ast Z_{0011 \ldots 0} + Z_{11000} \ast Z_{0101 \ldots 0} + Z_{11100} \ast Z_{0111 \ldots 0}, \\
        A_{101 \ldots 01} = Z_{10001} \ast Z_{0001 \ldots 0} + Z_{10101} \ast Z_{0011 \ldots 0} + Z_{11001} \ast Z_{0101 \ldots 0} + Z_{11101} \ast Z_{0111 \ldots 0}, \\
        A_{101 \ldots 10} = Z_{10000} \ast Z_{0001 \ldots 1} + Z_{10100} \ast Z_{0011 \ldots 1} + Z_{11000} \ast Z_{0101 \ldots 1} + Z_{11100} \ast Z_{0111 \ldots 1}, \\
        A_{101 \ldots 11} = Z_{10001} \ast Z_{0001 \ldots 1} + Z_{10101} \ast Z_{0011 \ldots 1} + Z_{11001} \ast Z_{0101 \ldots 1} + Z_{11101} \ast Z_{0111 \ldots 1}, \\
        \dots \\
        A_{110 \ldots 00} = Z_{10010} \ast Z_{0000 \ldots 0} + Z_{10110} \ast Z_{0010 \ldots 0} + Z_{11010} \ast Z_{0100 \ldots 0} + Z_{11110} \ast Z_{0110 \ldots 0}, \\
        A_{110 \ldots 01} = Z_{10011} \ast Z_{0000 \ldots 0} + Z_{10111} \ast Z_{0010 \ldots 0} + Z_{11011} \ast Z_{0100 \ldots 0} + Z_{11111} \ast Z_{0110 \ldots 0}, \\
        A_{110 \ldots 10} = Z_{10010} \ast Z_{0000 \ldots 1} + Z_{10110} \ast Z_{0010 \ldots 1} + Z_{11010} \ast Z_{0100 \ldots 1} + Z_{11110} \ast Z_{0110 \ldots 1}, \\
        A_{110 \ldots 11} = Z_{10011} \ast Z_{0000 \ldots 1} + Z_{10111} \ast Z_{0010 \ldots 1} + Z_{11011} \ast Z_{0100 \ldots 1} + Z_{11111} \ast Z_{0110 \ldots 1}, \\
        \dots \\
        A_{111 \ldots 00} = Z_{10010} \ast Z_{0001 \ldots 0} + Z_{10110} \ast Z_{0011 \ldots 0} + Z_{11010} \ast Z_{0101 \ldots 0} + Z_{11110} \ast Z_{0111 \ldots 0}, \\
        A_{111 \ldots 01} = Z_{10011} \ast Z_{0001 \ldots 0} + Z_{10111} \ast Z_{0011 \ldots 0} + Z_{11011} \ast Z_{0101 \ldots 0} + Z_{11111} \ast Z_{0111 \ldots 0}, \\
        A_{111 \ldots 10} = Z_{10010} \ast Z_{0001 \ldots 1} + Z_{10110} \ast Z_{0011 \ldots 1} + Z_{11010} \ast Z_{0101 \ldots 1} + Z_{11110} \ast Z_{0111 \ldots 1}, \\
        A_{111 \ldots 11} = Z_{10011} \ast Z_{0001 \ldots 1} + Z_{10111} \ast Z_{0011 \ldots 1} + Z_{11011} \ast Z_{0101 \ldots 1} + Z_{11111} \ast Z_{0111 \ldots 1}, \\
        \dots \quad .
    \end{aligned}
\end{equation}

Let us introduce column vectors $\overleftarrow{A_1}$ and  $\overleftarrow{Z_0}$ of dimension $2^{R+1}$ 
\begin{equation} \label{eq:20}
    \overleftarrow{A_1} = 
    \begin{bmatrix}
        A_{100 \ldots 00} \\
        A_{100 \ldots 01} \\
        A_{100 \ldots 10} \\
        A_{100 \ldots 11} \\
        \ldots \\
        \ldots \\
        A_{101 \ldots 00} \\
        A_{101 \ldots 01} \\
        A_{101 \ldots 10} \\
        A_{101 \ldots 11} \\
        \ldots \\
        \ldots \\
        A_{110 \ldots 00} \\
        A_{110 \ldots 01} \\
        A_{110 \ldots 10} \\
        A_{110 \ldots 11} \\
        \ldots \\
        \ldots \\
        A_{111 \ldots 00} \\
        A_{111 \ldots 01} \\
        A_{111 \ldots 10} \\
        A_{111 \ldots 11} \\
        \ldots \\
        \ldots
    \end{bmatrix}
    ,\qquad
    \overleftarrow{Z_0}=
    \begin{bmatrix}
        Z_{0000 \ldots 0} \\
        Z_{0000 \ldots 1} \\
        \ldots \\
        Z_{0001 \ldots 0} \\
        Z_{0001 \ldots 1} \\
        \ldots \\
        Z_{0010 \ldots 0} \\
        Z_{0010 \ldots 1} \\
        \ldots \\
        Z_{0011 \ldots 0} \\
        Z_{0011 \ldots 1} \\
        \ldots \\
        Z_{0100 \ldots 0} \\
        Z_{0100 \ldots 1} \\
        \ldots \\
        Z_{0101 \ldots 0} \\
        Z_{0101 \ldots 1} \\
        \ldots \\
        Z_{0110 \ldots 0} \\
        Z_{0110 \ldots 1} \\
        \ldots \\
        Z_{0111 \ldots 0} \\
        Z_{0111 \ldots 1} \\
        \ldots
    \end{bmatrix}
    .
\end{equation}

and four $2^R\times2^R$ matrices

\begin{equation} \label{eq:21}
\setcounter{MaxMatrixCols}{12}
    \begin{aligned}
    \overleftrightarrow{\zeta_{00}}
    & =
    \begin{bmatrix}
        Z_{10000} & 0 & \ldots & 0 & 0 & \ldots & Z_{10100} & 0 & \ldots & 0 & 0 & \ldots \\
        Z_{10001} & 0 & \ldots & 0 & 0 & \ldots & Z_{10101} & 0 & \ldots & 0 & 0 & \ldots \\
        0 & Z_{10000} & \ldots & 0 & 0 & \ldots & 0 & Z_{10100} & \ldots & 0 & 0 & \ldots \\
        0 & Z_{10001} & \ldots & 0 & 0 & \ldots & 0 & Z_{10101} & \ldots & 0 & 0 & \ldots \\
        \ldots & \ldots & \ldots & \ldots & \ldots & \ldots & \ldots & \ldots & \ldots & \ldots & \ldots & \ldots \\
        \ldots & \ldots & \ldots & \ldots & \ldots & \ldots & \ldots & \ldots & \ldots & \ldots & \ldots & \ldots \\
        0 & 0 & \ldots & Z_{10000} & 0 & \ldots & 0 & 0 & \ldots & Z_{10100} & 0 & \ldots \\
        0 & 0 & \ldots & Z_{10001} & 0 & \ldots & 0 & 0 & \ldots & Z_{10101} & 0 & \ldots \\
        0 & 0 & \ldots & 0 & Z_{10000} & \ldots & 0 & 0 & \ldots & 0 & Z_{10100} & \ldots \\
        0 & 0 & \ldots & 0 & Z_{10001} & \ldots & 0 & 0 & \ldots & 0 & Z_{10101} & \ldots \\
        \ldots & \ldots & \ldots & \ldots & \ldots & \ldots & \ldots & \ldots & \ldots & \ldots & \ldots & \ldots \\
        \ldots & \ldots & \ldots & \ldots & \ldots & \ldots & \ldots & \ldots & \ldots & \ldots & \ldots & \ldots     
    \end{bmatrix}
    , \\
    \\
     \overleftrightarrow{\zeta_{01}}
    & =
    \begin{bmatrix}
        Z_{11000} & 0 & \ldots & 0 & 0 & \ldots & Z_{11100} & 0 & \ldots & 0 & 0 & \ldots \\
        Z_{11001} & 0 & \ldots & 0 & 0 & \ldots & Z_{11101} & 0 & \ldots & 0 & 0 & \ldots \\
        0 & Z_{11000} & \ldots & 0 & 0 & \ldots & 0 & Z_{11100} & \ldots & 0 & 0 & \ldots \\
        0 & Z_{11001} & \ldots & 0 & 0 & \ldots & 0 & Z_{11101} & \ldots & 0 & 0 & \ldots \\
        \ldots & \ldots & \ldots & \ldots & \ldots & \ldots & \ldots & \ldots & \ldots & \ldots & \ldots & \ldots \\
        \ldots & \ldots & \ldots & \ldots & \ldots & \ldots & \ldots & \ldots & \ldots & \ldots & \ldots & \ldots \\
        0 & 0 & \ldots & Z_{11000} & 0 & \ldots & 0 & 0 & \ldots & Z_{11100} & 0 & \ldots \\
        0 & 0 & \ldots & Z_{11001} & 0 & \ldots & 0 & 0 & \ldots & Z_{11101} & 0 & \ldots \\
        0 & 0 & \ldots & 0 & Z_{11000} & \ldots & 0 & 0 & \ldots & 0 & Z_{11100} & \ldots \\
        0 & 0 & \ldots & 0 & Z_{11001} & \ldots & 0 & 0 & \ldots & 0 & Z_{11101} & \ldots \\
        \ldots & \ldots & \ldots & \ldots & \ldots & \ldots & \ldots & \ldots & \ldots & \ldots & \ldots & \ldots \\
        \ldots & \ldots & \ldots & \ldots & \ldots & \ldots & \ldots & \ldots & \ldots & \ldots & \ldots & \ldots     
    \end{bmatrix}
    , \\
    \\
    \overleftrightarrow{\zeta_{10}}
    & =
    \begin{bmatrix}
        Z_{10010} & 0 & \ldots & 0 & 0 & \ldots & Z_{10110} & 0 & \ldots & 0 & 0 & \ldots \\
        Z_{10011} & 0 & \ldots & 0 & 0 & \ldots & Z_{10111} & 0 & \ldots & 0 & 0 & \ldots \\
        0 & Z_{10010} & \ldots & 0 & 0 & \ldots & 0 & Z_{10110} & \ldots & 0 & 0 & \ldots \\
        0 & Z_{10011} & \ldots & 0 & 0 & \ldots & 0 & Z_{10111} & \ldots & 0 & 0 & \ldots \\
        \ldots & \ldots & \ldots & \ldots & \ldots & \ldots & \ldots & \ldots & \ldots & \ldots & \ldots & \ldots \\
        \ldots & \ldots & \ldots & \ldots & \ldots & \ldots & \ldots & \ldots & \ldots & \ldots & \ldots & \ldots \\
        0 & 0 & \ldots & Z_{10010} & 0 & \ldots & 0 & 0 & \ldots & Z_{10110} & 0 & \ldots \\
        0 & 0 & \ldots & Z_{10011} & 0 & \ldots & 0 & 0 & \ldots & Z_{10111} & 0 & \ldots \\
        0 & 0 & \ldots & 0 & Z_{10010} & \ldots & 0 & 0 & \ldots & 0 & Z_{10110} & \ldots \\
        0 & 0 & \ldots & 0 & Z_{10011} & \ldots & 0 & 0 & \ldots & 0 & Z_{10111} & \ldots \\
        \ldots & \ldots & \ldots & \ldots & \ldots & \ldots & \ldots & \ldots & \ldots & \ldots & \ldots & \ldots \\
        \ldots & \ldots & \ldots & \ldots & \ldots & \ldots & \ldots & \ldots & \ldots & \ldots & \ldots & \ldots     
    \end{bmatrix}
    , \\
    \\
    \overleftrightarrow{\zeta_{11}}
    & =
    \begin{bmatrix}
        Z_{11010} & 0 & \ldots & 0 & 0 & \ldots & Z_{11110} & 0 & \ldots & 0 & 0 & \ldots \\
        Z_{11011} & 0 & \ldots & 0 & 0 & \ldots & Z_{11111} & 0 & \ldots & 0 & 0 & \ldots \\
        0 & Z_{11010} & \ldots & 0 & 0 & \ldots & 0 & Z_{11110} & \ldots & 0 & 0 & \ldots \\
        0 & Z_{11011} & \ldots & 0 & 0 & \ldots & 0 & Z_{11111} & \ldots & 0 & 0 & \ldots \\
        \ldots & \ldots & \ldots & \ldots & \ldots & \ldots & \ldots & \ldots & \ldots & \ldots & \ldots & \ldots \\
        \ldots & \ldots & \ldots & \ldots & \ldots & \ldots & \ldots & \ldots & \ldots & \ldots & \ldots & \ldots \\
        0 & 0 & \ldots & Z_{11010} & 0 & \ldots & 0 & 0 & \ldots & Z_{11110} & 0 & \ldots \\
        0 & 0 & \ldots & Z_{11011} & 0 & \ldots & 0 & 0 & \ldots & Z_{11111} & 0 & \ldots \\
        0 & 0 & \ldots & 0 & Z_{11010} & \ldots & 0 & 0 & \ldots & 0 & Z_{11110} & \ldots \\
        0 & 0 & \ldots & 0 & Z_{11011} & \ldots & 0 & 0 & \ldots & 0 & Z_{11111} & \ldots \\
        \ldots & \ldots & \ldots & \ldots & \ldots & \ldots & \ldots & \ldots & \ldots & \ldots & \ldots & \ldots \\
        \ldots & \ldots & \ldots & \ldots & \ldots & \ldots & \ldots & \ldots & \ldots & \ldots & \ldots & \ldots     
    \end{bmatrix}
    .
    \end{aligned}
\end{equation}

Then we get (\ref{eq:14})
\begin{equation*}
    \overleftarrow{A_1} = \overleftrightarrow{Z_1} * \overleftarrow{Z_0},
\end{equation*}

where $2^{R+1}\times2^{R+1}$ matrix $\overleftrightarrow{Z_1}$ consists of four $2^R\times2^R$ block matrices

\begin{equation} \label{eq:22}
    \overleftrightarrow{Z_1}
    =
    \begin{pmatrix}
        \overleftrightarrow{\zeta_{00}} & \overleftrightarrow{\zeta_{01}} \\
        \overleftrightarrow{\zeta_{10}} & \overleftrightarrow{\zeta_{11}}
    \end{pmatrix}
    .
\end{equation}

And so on. In the end, we introduce the row vector for the finish cell  $\overrightarrow{Z_{N+1}}$  and get the matrix model (\ref{eq:9}) 

\begin{equation*}
    Z=\overrightarrow{Z_{N+1}} * \overleftrightarrow{Z_N} * \ldots *\overleftrightarrow{Z_n} * \ldots * \overleftrightarrow{Z_2} * \overleftrightarrow{Z_1}* \overleftarrow{Z_0}=\overrightarrow{Z_{N+1}}* \left( \prod_{n=N}^1 \overleftrightarrow{Z_n} \right) * \overleftarrow{Z_0}.
\end{equation*}

\subsection{2D cyclic shift matrices}

The structure of four $2^R\times2^R$ matrices $\overleftrightarrow{\zeta_{00}}, \  \overleftrightarrow{\zeta_{01}}, \ \overleftrightarrow{\zeta_{10}}, \ \overleftrightarrow{\zeta_{11}}$ is similar to the structure of the matrix $\overleftrightarrow{Z_1}$ from (\ref{eq:13}). As in subsection \ref{subsec:SimplestCyclicShiftMatrices}, we introduce $2^R\times2^R$ left cyclic shift matrix $\overleftrightarrow{P_l}$ and its inverse as right cyclic shift matrix $\overleftrightarrow{P_r}=\overleftrightarrow{P_l}^{-1}$.

\begin{equation} \label{eq:23}
    \begin{aligned}
        \overleftrightarrow{P_l}
        & =
        \begin{pmatrix}
            1 & 0 & 0 & 0 & \ldots & \ldots & 0 & 0 & 0 & 0 & \ldots & \ldots \\
            0 & 0 & 1 & 0 & \ldots & \ldots & 0 & 0 & 0 & 0 & \ldots & \ldots \\
            \ldots & \ldots & \ldots & \ldots & \ldots & \ldots & \ldots & \ldots & \ldots & \ldots & \ldots & \ldots \\
            0 & 0 & 0 & 0 & \ldots & \ldots & 1 & 0 & 0 & 0 & \ldots & \ldots \\
            0 & 0 & 0 & 0 & \ldots & \ldots & 0 & 0 & 1 & 0 & \ldots & \ldots \\
            \ldots & \ldots & \ldots & \ldots & \ldots & \ldots & \ldots & \ldots & \ldots & \ldots & \ldots & \ldots \\
            0 & 1 & 0 & 0 & \ldots & \ldots & 0 & 0 & 0 & 0 & \ldots & \ldots \\
            0 & 0 & 0 & 1 & \ldots & \ldots & 0 & 0 & 0 & 0 & \ldots & \ldots \\
            \ldots & \ldots & \ldots & \ldots & \ldots & \ldots & \ldots & \ldots & \ldots & \ldots & \ldots & \ldots \\
            0 & 0 & 0 & 0 & \ldots & \ldots & 0 & 1 & 0 & 0 & \ldots & \ldots \\
            0 & 0 & 0 & 0 & \ldots & \ldots & 0 & 0 & 0 & 1 & \ldots & \ldots \\
            \ldots & \ldots & \ldots & \ldots & \ldots & \ldots & \ldots & \ldots & \ldots & \ldots & \ldots & \ldots
        \end{pmatrix}
        , \\
        \\
        \overleftrightarrow{P_r}
        & =
        \begin{pmatrix}
            1 & 0 & \ldots & 0 & 0 & \ldots & 0 & 0 & \ldots & 0 & 0 & \ldots \\
            0 & 0 & \ldots & 0 & 0 & \ldots & 1 & 0 & \ldots & 0 & 0 & \ldots \\
            0 & 1 & \ldots & 0 & 0 & \ldots & 0 & 0 & \ldots & 0 & 0 & \ldots \\
            0 & 0 & \ldots & 0 & 0 & \ldots & 0 & 1 & \ldots & 0 & 0 & \ldots \\
            \ldots & \ldots & \ldots & \ldots & \ldots & \ldots & \ldots & \ldots & \ldots & \ldots & \ldots & \ldots \\
            \ldots & \ldots & \ldots & \ldots & \ldots & \ldots & \ldots & \ldots & \ldots & \ldots & \ldots & \ldots \\
            0 & 0 & \ldots & 1 & 0 & \ldots & 0 & 0 & \ldots & 0 & 0 & \ldots \\
            0 & 0 & \ldots & 0 & 0 & \ldots & 0 & 0 & \ldots & 1 & 0 & \ldots \\
            0 & 0 & \ldots & 0 & 1 & \ldots & 0 & 0 & \ldots & 0 & 0 & \ldots \\
            0 & 0 & \ldots & 0 & 0 & \ldots & 0 & 0 & \ldots & 0 & 1 & \ldots \\
            \ldots & \ldots & \ldots & \ldots & \ldots & \ldots & \ldots & \ldots & \ldots & \ldots & \ldots & \ldots \\
            \ldots & \ldots & \ldots & \ldots & \ldots & \ldots & \ldots & \ldots & \ldots & \ldots & \ldots & \ldots
        \end{pmatrix}
        .
    \end{aligned}
\end{equation}

Then the matrix $Z_1$ from (\ref{eq:22}) can be represented as
\begin{equation} \label{eq:24}
    \overleftrightarrow{Z_1}
    =
    \begin{pmatrix}
        \overleftrightarrow{B_{00}} & \overleftrightarrow{B_{01}} \\
        \overleftrightarrow{B_{10}} & \overleftrightarrow{B_{11}}
    \end{pmatrix}
    *
    \begin{pmatrix}
        \overleftrightarrow{P_r} & 0 \\
        0 & \overleftrightarrow{P_r}
    \end{pmatrix}
    ,
\end{equation}

where four $2^R\times2^R$ block matrices $\overleftrightarrow{B_{00}}, \ \overleftrightarrow{B_{01}}, \ \overleftrightarrow{B_{10}}, \ \overleftrightarrow{B_{11}}$ are block-diagonal, along the diagonal of which there are identical $2\times2$ blocks. 

Consider elements of these four $2^R\times2^R$ block matrices. An element with compound row number  $i_1 i_2 \ldots i_R$ and compound column number  $j_1 j_2 \ldots j_R$ is non-zero only if all row sub-numbers except the last sub-number $R$ are equal to the corresponding column sub-numbers. To emphasize this, we denote these four block matrices as $2\times2$ matrices with the number $[R]$. Then four block matrices from (\ref{eq:24}) are 

\begin{equation} \label{eq:25}
    \begin{aligned}
    \overleftrightarrow{B_{00}}
    & =
    \overleftrightarrow{\zeta_{00}} * \overleftrightarrow{P_l}
    =
    \begin{pmatrix}
        Z_{10000} & Z_{10100} \\
        Z_{10001} & Z_{10101}
    \end{pmatrix}
    _{[R]}
    =
    \begin{pmatrix}
        Z_{10000} & Z_{10100} & 0 & 0 & \ldots & 0 & 0 \\
        Z_{10001} & Z_{10101} & 0 & 0 & \ldots & 0 & 0 \\
        0 & 0 & Z_{10000} & Z_{10100} & \ldots & 0 & 0 \\
        0 & 0 & Z_{10001} & Z_{10101} & \ldots & 0 & 0 \\
        \ldots & \ldots & \ldots & \ldots & \ldots & \ldots & \ldots \\
        0 & 0 & 0 & 0 & \ldots & Z_{10000} & Z_{10100} \\
        0 & 0 & 0 & 0 & \ldots & Z_{10001} & Z_{10101}
    \end{pmatrix}
    , \\
    \\
    \overleftrightarrow{B_{01}}
    & = 
    \overleftrightarrow{\zeta_{01}} * \overleftrightarrow{P_l}
    =
    \begin{pmatrix}
        Z_{11000} & Z_{11100} \\
        Z_{11001} & Z_{11101}
    \end{pmatrix}
    _{[R]}
    =
    \begin{pmatrix}
        Z_{11000} & Z_{11100} & 0 & 0 & \ldots & 0 & 0 \\
        Z_{11001} & Z_{11101} & 0 & 0 & \ldots & 0 & 0 \\
        0 & 0 & Z_{11000} & Z_{11100} & \ldots & 0 & 0 \\
        0 & 0 & Z_{11001} & Z_{11101} & \ldots & 0 & 0 \\
        \ldots & \ldots & \ldots & \ldots & \ldots & \ldots & \ldots \\
        0 & 0 & 0 & 0 & \ldots & Z_{11000} & Z_{11100} \\
        0 & 0 & 0 & 0 & \ldots & Z_{11001} & Z_{11101}
    \end{pmatrix}
    , \\
    \\
    \overleftrightarrow{B_{10}}
    & =
    \overleftrightarrow{\zeta_{10}} * \overleftrightarrow{P_l}
    =
    \begin{pmatrix}
        Z_{10010} & Z_{10110} \\
        Z_{10011} & Z_{10111}
    \end{pmatrix}
    _{[R]}
    =
    \begin{pmatrix}
        Z_{10010} & Z_{10110} & 0 & 0 & \ldots & 0 & 0 \\
        Z_{10011} & Z_{10111} & 0 & 0 & \ldots & 0 & 0 \\
        0 & 0 & Z_{10010} & Z_{10110} & \ldots & 0 & 0 \\
        0 & 0 & Z_{10011} & Z_{10111} & \ldots & 0 & 0 \\
        \ldots & \ldots & \ldots & \ldots & \ldots & \ldots & \ldots \\
        0 & 0 & 0 & 0 & \ldots & Z_{10010} & Z_{10110} \\
        0 & 0 & 0 & 0 & \ldots & Z_{10011} & Z_{10111}
    \end{pmatrix}
    , \\
    \\
    \overleftrightarrow{B_{11}}
    & =
    \overleftrightarrow{\zeta_{11}} * \overleftrightarrow{P_l}
    =
    \begin{pmatrix}
        Z_{11010} & Z_{11110} \\
        Z_{11011} & Z_{11111}
    \end{pmatrix}
    _{[R]}
    =
    \begin{pmatrix}
        Z_{11010} & Z_{11110} & 0 & 0 & \ldots & 0 & 0 \\
        Z_{11011} & Z_{11111} & 0 & 0 & \ldots & 0 & 0 \\
        0 & 0 & Z_{11010} & Z_{11110} & \ldots & 0 & 0 \\
        0 & 0 & Z_{11011} & Z_{11111} & \ldots & 0 & 0 \\
        \ldots & \ldots & \ldots & \ldots & \ldots & \ldots & \ldots \\
        0 & 0 & 0 & 0 & \ldots & Z_{11010} & Z_{11110} \\
        0 & 0 & 0 & 0 & \ldots & Z_{11011} & Z_{11111}
    \end{pmatrix}
    .
    \end{aligned}
\end{equation}

Let us indicate the further used properties of $2^R \times 2^R$ shift matrices $\overleftrightarrow{P_l}$ and $\overleftrightarrow{P_r}$  

\begin{equation} \label{eq:26}
    \overleftrightarrow{P_l}^R
    =
    \overleftrightarrow{P_r}^R
    =
    \overleftrightarrow{1}
    ,
\end{equation}

where $\overleftrightarrow{1}$ is $2^R \times 2^R$ identity matrix.

Moreover, for arbitrary matrix $\overleftrightarrow{B_{[r]}}$ the equalities hold

\begin{equation} \label{eq:27}
    \overleftrightarrow{P_r}
    *
    \overleftrightarrow{B}_{[r]}
    *
    \overleftrightarrow{P_l}
    =
    \overleftrightarrow{B}_{[r+1]}
    ,\ 
    \overleftrightarrow{P_l}
    *
    \overleftrightarrow{B}_{[r]}
    *
    \overleftrightarrow{P_r}
    =
    \overleftrightarrow{B}_{[r-1]}
    .
\end{equation}

\section{Conclusion}

\begin{enumerate}
    \item In this paper, Section 2 describes the simple method for constructing the exact matrix model for the generalized Ising model.
    \item Using this method, the exact matrix model is constructed for two-dimensional system (see Figure \ref{fig:fig1}). It is (\ref{eq:24}) taking into account (\ref{eq:23}), (\ref{eq:25}) and (\ref{eq:17}).
\end{enumerate}

\bibliographystyle{unsrtnat}
\bibliography{references}

\end{document}